\documentclass[lettersize,journal]{IEEEtran}
\usepackage[table]{xcolor}
\usepackage{amsmath,amsfonts}
\usepackage{algorithmic}
\usepackage{algorithm}
\usepackage{array}
\usepackage[caption=false,font=normalsize,labelfont=sf,textfont=sf]{subfig}
\usepackage{bm}
\usepackage{textcomp}
\usepackage{stfloats}
\usepackage{url}
\usepackage{verbatim}
\usepackage{graphicx}
\usepackage{cite}
\usepackage{amsmath}
\usepackage{multirow}
\usepackage{makecell}
\usepackage{enumitem}
\usepackage{amsmath}
\usepackage{soul} % 导入 soul 包
\usepackage{color, xcolor}
\sethlcolor{cyan}
\begin{document}

\title{Privacy-Preserving Electricity Theft Detection based on Blockchain} 

\author{Zhiqiang Zhao, Yining Liu,~\IEEEmembership{Member~IEEE,} Zhixin Zeng, Zhixiong Chen, Huiyu Zhou
        % <-this % stops a space
%\thanks{This paper was produced by the IEEE Publication Technology Group. They are in Piscataway, NJ.}% <-this % stops a space
\thanks{Manuscript received July 13, 2022; revised September 14, 2022 and December 27, 2022; accepted February 5, 2023. This work was supported by Natural Science Foundation of China (No. 62072133) and the Fujian Key Laboratory of Financial Information Processing (Putian University) (No. JXC202201). (Corresponding author: Yining Liu.)}
\thanks{Zhiqiang Zhao, Yining Liu and Zhixin Zeng are with
School of Computer Science and Information Security, Guilin University of Electronic Technology, Guilin 541004, China (e-mail: lanren0909@gmail.com, lyn7311@sina.com and bestzengzx@gmail.com).}
\thanks{Zhixiong Chen is with the Fujian Key Laboratory of Financial Information Processing and the Key Laboratory of Applied Mathematics of Fujian Province University, Putian University, Putian, Fujian 351100, China (e-mail: ptczx@126.com).}
\thanks{Huiyu Zhou is with the School of Computing and Mathematical
Sciences, University of Leicester, LE1 7RH Leicester, U.K. (e-mail: hz143@leicester.ac.uk).}
%\thanks{Color versions of one or more of the figures in this paper are available online at http://ieeexplore.ieee.org.}
%\thanks{Digital Object Identifier *}
}

% The paper headers
\markboth{IEEE TRANSACTIONS ON SMART GRID}%
{Shell \MakeLowercase{\textit{et al.}}: }

%\IEEEpubid{0000--0000/00\$00.00~\copyright~2021 IEEE}
% Remember, if you use this you must call \IEEEpubidadjcol in the second
% column for its text to clear the IEEEpubid mark.

\maketitle

\begin{abstract}
In most electricity theft detection schemes, consumers' power consumption data is directly input into the detection center. Although it is valid in detecting the theft of consumers, the privacy of all consumers is at risk unless the detection center is assumed to be trusted. In fact, it is impractical. Moreover, existing schemes may result in some security problems, such as the collusion attack due to the presence of a trusted third party, and malicious data tampering caused by the system operator (SO) being attacked. Aiming at the problems above, we propose a blockchain-based privacy-preserving electricity theft detection scheme without a third party. Specifically, the proposed scheme uses an improved functional encryption scheme to enable electricity theft detection and load monitoring while preserving consumers' privacy; distributed storage of consumers' data with blockchain to resolve security problems such as data tampering, etc. Meanwhile, we build a long short-term memory network (LSTM) model to perform higher accuracy for electricity theft detection. The proposed scheme is evaluated in a real environment, and the results show that it is more accurate in electricity theft detection within acceptable communication and computational overhead. Our system analysis demonstrates that the proposed scheme can resist various security attacks and preserve consumers’ privacy.

\end{abstract}


\section{Introduction}
\IEEEPARstart{S}{mart} grid (SG) is an advanced grid integrating smart technology, which uses smart meters (SMs) to collect, analyze and process fine-grained power consumption data from consumers to manage energy effectively \cite{gungor2012survey}. While the smart grid brings convenience, it also brings serious challenges\cite{wang2018review}. For one thing, the communication of the smart grid is exposed to potential malicious attacks, such as data tampering attack and false data injection. If these malicious attacks cannot be resisted, the smart grid will be unable to operate normally \cite{zeng2022msda}. For another thing, electricity theft has become a widespread phenomenon in the smart grid. Annual economic losses due to electricity theft are estimated to be about 170 million dollars in the United Kingdom \cite{xia2019sai} and 6 billion dollars in the United States\cite{mcdaniel2009security}. Meanwhile, electricity theft can also seriously affect energy management and endanger the normal operation of the smart grid \cite{gope2018privacy}.

Since the smart grid has access to consumers' fine-grained power consumption data, the traditional machine learning model\cite{jokar2015electricity} and deep learning model \cite{ref9} based on big data have achieved good performance. However, directly giving fine-grained power consumption data of consumers to the SO raises serious privacy issues \cite{rubio2017recommender}. Meanwhile, as the security and privacy of data are becoming more and more concerned, related laws and regulations have been proposed, such as the General Data Protection Regulations (GDPR) in Europe, and the utilities' disregard for privacy aspects could lead to strong consumer objection and significant curtailment of service deployment\cite{hoenkamp2011neglected}. Therefore, there is an urgent demand for a privacy-preserving electricity theft detection scheme.

Although existing schemes are beginning to consider the privacy of consumers' power consumption data during the electricity theft detection process, most schemes have serious challenges. On the one hand, a serious threat is the potential leakage of consumers' privacy due to the presence of a trusted third party. In \cite{yao2019energy}, the model requires a fully trusted third party to perform the detection using the original consumers' data. However, it is difficult to guarantee that the third party is fully trustworthy in reality, so consumers' privacy is still at risk of being compromised. In \cite{ibrahem2020efficient}, the scheme requires a fully trusted key distribution center. However, once the SO colludes with the key distribution center, then the SO can get the consumers' raw power consumption data, which leads to consumers' privacy leakage. On the other hand, the security of data and smart grid is not considered. In \cite{yao2019energy}, if the trustworthy detection center is maliciously attacked, then it will possibly lead to malicious data tampering. In \cite{ibrahem2020efficient}, this scheme does not verify the legitimacy of the transmitted data, so it is unable to resist data tampering and forgery attacks. The existing schemes do not consider the security of the smart grid in operation and data tampering due to centralized data storage when performing electricity theft detection, thus making it impossible to achieve electricity theft detection. Therefore, how to accomplish the security of smart grid operations and consumers' privacy while utilizing consumers' power consumption data is a major challenge of current research.

In this paper, we aim to achieve more secure electricity theft detection and load
monitoring without the involvement of a third party. The main contributions of this work are threefold:

\begin{enumerate}
\item We propose a blockchain-based electricity theft detection scheme, which uses the distributed storage of blockchain to solve security problems such as data tampering of centralized storage, etc.
\item We improve the functional encryption scheme\cite{ibrahem2020efficient} to enable privacy-preserving electricity theft detection and load monitoring without a trusted key distribution center, which eliminates potential security and privacy problems caused by a third party.
\item We build an electricity theft detection model based on long short-term memory networks that are more suitable for processing time-series data, and the model parameter settings are analyzed to obtain higher performance.

\end{enumerate}

\begin{table*}[!t]
\renewcommand\arraystretch{1.4}
\caption{The comparison of related work\label{tab:table1}}
\centering
\begin{tabular}{|c|c|c|c|c|c|c|}
\rowcolor{gray!45} 
\hline 
& \textbf{Joker et al.\cite{jokar2015electricity}}& \textbf{Wen et al.\cite{wen2021feddetect}} & \textbf{Yao et al.\cite{yao2019energy}} & \textbf{I.Ibrahem et al.\cite{ibrahem2020efficient}} & \textbf{Richardson et al.\cite{richardson2016privacy}} & \textbf{Nabil et al.\cite{nabil2019ppetd}}  \\
\hline
\textbf{Technique adopted}& SVM & Federal Learning & CNN & FNN & DBSCAN & 1-D CNN \\
\hline
\textbf{Attack Defense} & No & No & No & No & No & No  \\
\hline
\textbf{Grid monitoring} &Yes & No & Yes & Yes &No & Yes  \\
\hline
\textbf{Third-party}& No &Yes & Yes & Yes  & No & No  \\
\hline

\end{tabular}

\end{table*}
The remainder of this paper is organized as follows. In Section II, we review the related work. Section III illustrates the related knowledge. In Section IV, we define the system model and design goals. Section V presents the proposed scheme. Experimental results and system characterization are presented in Sections VI and VII, respectively. Finally, the paper is summarized in Section VIII.

\section{Related Work}
In this section, we briefly review recent research work on electricity theft detection schemes in the smart grid and the distributed blockchain-based smart grid framework.

Currently, due to the seriousness of the electricity theft problem and the importance of privacy-preserving, we focus on electricity theft detection schemes with privacy-preserving, which can be broadly classified into two categories with or without the participation of a third party. The comparison of the related work is given in Table I.

In the case of schemes where a third party is involved, the third party is used to perform tasks such as distributing keys and performing electricity theft detection, etc. Wen et al.\cite{wen2021feddetect} proposed a privacy-preserving federal learning framework consisting of a data center, a control center, and multiple detection stations, which requires a high cost to complete the system. Moreover, the authors' scheme does not consider other functional requirements such as load monitoring. Yao et al.\cite{yao2019energy} proposed to send the encrypted data of SMs to a fully trusted detection center to decrypt and then detect using the convolutional Neural Network (CNN) model, meanwhile, SMs send the encrypted data to an untrusted center that aggregates power consumption data for load monitoring. In \cite{ibrahem2020efficient}, Ibrahem et al. proposed to use functional encryption and the feed-forward neural network (FNN) to perform electricity theft detection and privacy protection under the condition that the key distribution center is fully trusted. All of the above schemes assume that the third party is trustworthy, but in practice, consumers' privacy can still be compromised such as once the third party colludes with other entities. Untrustworthy third parties have caused the above-mentioned problems in other areas as well\cite{fu2020towards}, so it is important to eliminate the risks associated with the presence of untrustworthy third parties.

In schemes where no third party is required, the scheme is performed by only two entities, SM and SO. Joker et al.\cite{jokar2015electricity} proposed to use the support vector machine (SVM) to monitor consumption pattern anomalies and identify suspicious consumers in the case of low sampling of consumers' power consumption data. However, this scheme is difficult to resist malicious attacks, such as replay attacks, fake data injection, etc. In \cite{richardson2016privacy}, the Euclidean distance between the normalized photovoltaic power output of any two installations in the region in a day is calculated by homomorphic encryption. Then the Euclidean distances are clustered to analyze the anomalous users. However, this scheme detects energy theft from the perspective of energy output, and when a smart meter is tampered with due to external attacks, it can no longer be detected properly. Meanwhile, the authors' scheme cannot obtain the sum of power consumption in the region for load monitoring. Nabil et al. \cite{nabil2019ppetd} proposed a CNN machine learning model based on secure two-party computation protocols using arithmetic and binary circuits. This scheme requires high computation and communication overhead to complete the detection of a consumer, which takes at least 35 minutes for detection and a minimum of 1375 MB for communication overhead. None of the above schemes consider the problem of ensuring the operational security of the smart grid when performing electricity theft detection, such as the data tampering problem when SO is maliciously attacked. Therefore, a more secure detection model with acceptable computation and communication overhead is needed.

To ensure the security of the smart grid, in \cite{liang2018distributed}, Liang et al. proposed a new distributed blockchain-based protection framework to enhance the self-defense of the modern power system. In \cite{song2021blockchain}, the authors designed a blockchain-based platform to prevent user data from being tampered with and proposed a multifaceted mechanism to protect user privacy. In \cite{hamouda2020novel}, Hamouda et al. proposed a blockchain-based comprehensive transactive energy market framework that enables a safer and fairer electricity market. Fan et al.\cite{fan2020decentralized} proposed a decentralized privacy-preserving data aggregation scheme for the smart grid based on blockchain, which uses the Paillier cryptographic algorithm to aggregate consumers' power consumption data. In \cite{hamouda2021centralized}, the authors proposed a new blockchain-based strategy for inter-connected microgrids energy trading that enhances the security and transparency of the platform. In\cite{zhang2022efficient}, the authors proposed an efficient and robust blockchain-based multidimensional data aggregation scheme in the smart grid to resist more internal and external attacks. Chen et al. \cite{chen2021trusted} proposed a blockchain-based framework to prevent energy market failures caused by dishonest participants.

There are many recent studies that consider the distributed blockchain-based smart grid framework can secure the grid. Meanwhile, it is also a good solution for electricity theft detection, and to advance the state of the art, we propose a blockchain-based privacy-preserving electricity theft detection scheme, which will be further explained and evaluated in the following sections.

\section{Preliminaries}

\subsection{Secure Aggregation}
Bonawitz et al. \cite{bonawitz2017practical} proposed a secure aggregation scheme where the server can only see the gradient after the aggregation is completed and cannot know the private true gradient value of each user. Unlike the original text, the proposed scheme uses the elliptic curve Diffie-Hellman key agreement. The steps of secure aggregation are as follows:

\subsubsection{Key agreement between arbitrary SMs}
Each ${SM_i}$ negotiates key masks with each other.
\begin{itemize}
\item KA.Setup$(\lambda )\to(E,G,g,p,q,H)$: The setup algorithm takes as input the security parameters $\lambda $. Then it outputs cyclic additive group $G$ of prime order $q$, a basis point $g$, a hash function $H$, and an elliptic curve $E$ on $GF(p)$ as well as a large prime number $p$. 
\item KA.Gen$(E,G,g,p,q,H)\to(x,xg)$: Each user chooses a random $x \in {Z_q}$ as own secret key $s_u^{sk}$ and calculates $xg$ as the public key $s_u^{pk}$.
\item KA.Agree$({x_u},{{x_v}g})\to{s_{u,v}}$: After receiving the public key ${x_v}g$ from user $v$, user $u$ uses its own secret key ${x_u}$ to generate ${s_{u,v}} = H({x_u}({x_v}g))$.
\end{itemize}
\subsubsection{Generating masks for aggregation} 

A mask is generated by key agreement between arbitrary users. Assume that all users form a user set   $\mathcal{U}$ in order and each user $u \in \mathcal{U}$ computes:
\begin{equation} 
{y_u} = {x_u} + \sum\limits_{v \in \mathcal{U},u < v} {{s_{u,v}}}  - \sum\limits_{v \in \mathcal{U},u > v} {{s_{v,u}}} 
\end{equation} 
where $u < v$ represents the users whose serial number is less than $v$, by the same token, we can get $u > v$.

Each user $u$ sends ${y_u}$ to the server, the server computes Eq. (2) to securely aggregate the secret keys.
\begin{equation} 
  \begin{aligned}
  z &= \sum\limits_{u \in \mathcal{U}} {{y_u}}  \hfill \\
    &= \sum\limits_{u \in \mathcal{U}} {({x_u} + \sum\limits_{v \in \mathcal{U},u < v} {{s_{u,v}}}  - \sum\limits_{v \in \mathcal{U},u > v} {{s_{v,u}}} )}  \hfill \\
    &= \sum\limits_{u \in \mathcal{U}} {{x_u}}  \hfill \\ 
  \end{aligned} 
\end{equation}

\subsection{Boneh-Lynn-Shacham Short Signature}
Boneh-Lynn-Shacham (BLS) short signature \cite{boneh2004short} is a signature algorithm that enables signature aggregation and speeds up block verification, which is divided into three phases: key generation, signature, and veriﬁcation.
\begin{enumerate}
\item Key generation: Sampling random number $x \in Z_q^ * $ as the private key and calculating the public key $PK = x \cdot g$.
\item Signature: The message $m$ is mapped to a point $H(m)$ in the cyclic group ${G_1}$. Generating signature $\delta    = x \cdot H(m)$.
\item Veriﬁcation: If $e(\delta  ,g) = e(H(m),PK)$, where $e:{G_1} \times {G_1} \to {G_2}$ is a bilinear map, then the signature is veriﬁed. Otherwise fails.
\end{enumerate}

\begin{figure}[!t]
\centering
\includegraphics[width=3.4in]{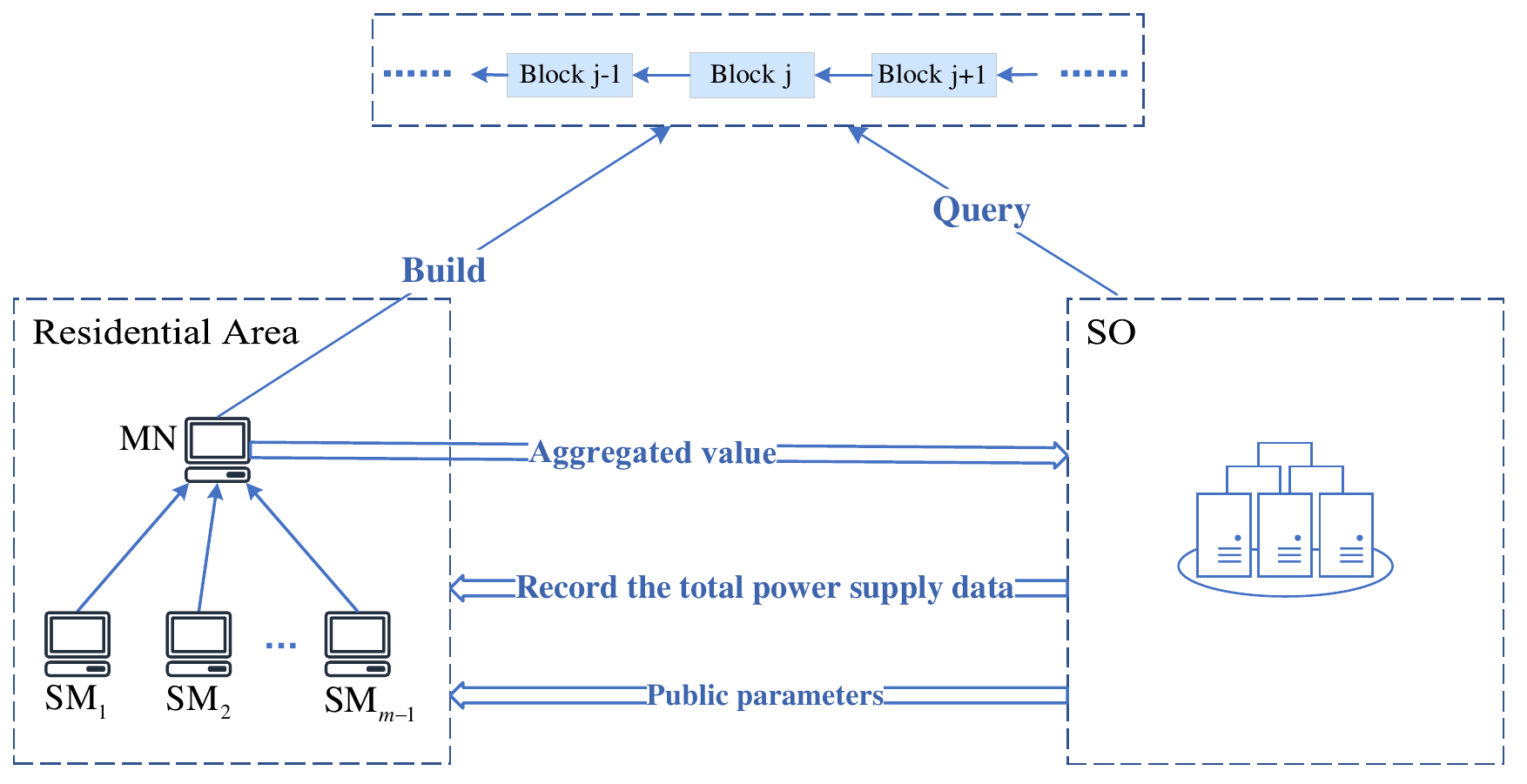}
\caption{System Model.}
\end{figure}

\section{System Model and Design Goals}
This section focuses on the construction of the system model and threat model as well as describes our design goals.

\subsection{System Model}
As shown in Fig. 1, the model of our system scheme includes smart meters in the residential area (RA), a system operator, and distribution transformer meters (DTMs). The function of each entity is described below.
\begin{enumerate}
\item Smart meter: SM is an electricity meter that sends the consumer's power consumption data to the mining node (MN) periodically (e.g., every 30 minutes) after implementing a predefined privacy-preserving scheme.
\item Mining node: The MN is a smart meter selected by the votes of all SMs in each residential area, it is responsible for verifying the legitimacy of the data, aggregating the encrypted data reported by SMs, and creating blocks to record power consumption data. If the MN goes down, all SMs will continue to vote for a new MN. If a malicious SM wants to become an MN, it needs to control at least $51\% $ of the SMs in the entire network to be elected as MN, but this is unrealistic.
\item System operator: The SO can generate system parameters and read the consumers' encrypted power consumption data through blockchain as well as get the real-time total power consumption $\sum\nolimits_i {{E_{S{M_i}}}(t)}$ of the area sent by MN, which are used for power consumption analysis and energy management. The SO uses a distribution transformer meter to record the total power supply data ${E_{DTM}}(T)$ for the residential area during the electricity theft detection period in order to judge the existence of electricity theft and perform electricity theft detection.

\end{enumerate}

\subsection{Threat Model}
For the system model proposed in the previous sub-section, we consider the threat from three aspects: consumers, the SO, and external attackers.
\begin{enumerate}
\item Consumers: Malicious consumers may falsify their power consumption data to reduce their bills. Also, they may collude with other consumers or SO to infer sensitive information about the victimized consumers. In addition, malicious consumers may deny their transmitted data when they are detected. With respect to MN, it may maliciously tamper the data reported by SMs.
\item SO: The SO is assumed to be honest but curious, i.e., it performs operations according to the protocol, but it may attempt to obtain fine-grained power consumption data from consumers to analyze valuable information.
\item External attackers: External attackers may attempt to eavesdrop on consumer communications to obtain consumer data, and may also forge malicious data to harm the SO, as well as initiate attacks on the SO to tamper with stored data.
\end{enumerate}

Therefore, the scheme aims to achieve the smart grid can resist malicious attacks and preserve consumers' privacy while still enabling energy management and electricity theft detection.

\subsection{Design Goals}
In order to protect the security and privacy of consumers' data without relying on a third party, the proposed scheme should achieve the following design goals:
\begin{enumerate}
\item Privacy preservation: For any one consumer, their original power consumption data is not obtainable by DTM, SO, and other consumers. Meanwhile, no entity can infer any private information from the encrypted data.
\item Confidentiality: Consumers' data is encrypted for transmission, storage, aggregation, and theft detection so that the original consumers' data cannot be recovered even if entities collude with each other.
\item Data unforgeability and non-repudiation: The consumers' encrypted data is signed and then transmitted to ensure that the data cannot be forged, while the transmission information is recorded in the blockchain to achieve data non-repudiation and data unforgeability.
\item Resist collusive attacks: The proposed scheme can resist the attack that smart grid entities collude with each other to obtain consumers' power consumption data.

\end{enumerate}

\section{The Proposed Scheme}
Our scheme consists of five phases: (1) system initialization phase; (2) reporting phase; (3) aggregation phase; (4) judgement phase; (5) electricity theft detection phase. The notations are listed in Table II. 
\subsection{Overview}
The main process of our scheme is summarized as follows:
\begin{itemize}
\item In the initialization phase, SO divides the residential area and generates the system parameters as well as parameters of the first layer of the neural network. The SMs in each detection region select the MN by Byzantine fault-tolerant consensus mechanism\cite{moniz2012byzantine}, while the SM generates encryption and decryption keys.
\item In the reporting phase, each SM encrypts the power consumption data $r(t)$ during the detection period $T = \{ {t_1},{t_{2,}} \cdots {t_d}\}$, then signs and sends encrypted data to the MN.
\item In the aggregation phase, MN verifies the legitimacy of the data, then constructs blocks and aggregates the power consumption data through the Merkle tree.
\item In the judgement phase, SO judges whether there is electricity theft in a region based on the difference between the DTM statistics and the aggregated data of MN within the tolerance range.
\item In the electricity theft detection phase, SO reads the encrypted data from the blockchain that is reported by each SM in the suspected electricity theft area during the theft detection period. The encrypted data are decrypted (still in ciphertext state after decryption) and then fed to the detection model to identify the electricity theft consumers.
\end{itemize}

\begin{table}[!t]
\caption{Notations\label{tab:table2}}
\centering
\renewcommand\arraystretch{1.2}
\begin{tabular}{ccc}
\hline
Notation & Description\\
\hline
${E_{DTM}}(t)$ & Power supply data for a residential area\\
$\sum\nolimits_i {{E_{S{M_i}}}(t)}$ & Total of uploaded data for all SMs \\
${\rm{S}}{{\rm{M}}_i}$ & i-th smart meter\\
${\mathbb{C}_i}[t]$ & Encrypted reading of $S{M_i}$ at time $t$\\
$\mathcal{U}$ & The set of SMs in the detection region\\
$W$ & The first layer’s weights of the model\\
$T$ & Electricity theft detection period\\
$TS_t^i$ & Timestamp of $S{M_i}$\\
${\delta _i}$ & Signature of $S{M_i}$\\
$DA$ & A decryption key for aggregating readings\\
$D{W_i}$ & Decryption keys for electricity theft detection\\
$|RA|$& Number of residential areas \\
$m$& Number of smart meters in the detection area \\
$d$& Number of readings for electricity theft detection period \\
\hline
\end{tabular}
\end{table}

\subsection{System Initialization}
System initialization includes three parts. First, SO generates the parameters of the system and the first layer’s weights of the model, and delineates $|RA|$ residential areas with $m$ SMs in each detection area. Second, all SMs in the region reach consensus to choose the MN. Third, Each SM generates its own keys.

\subsubsection{System parameters generation}  
\begin{itemize}
\item \textbf{Step 1:} The SO generate $(q,g,G,{G_1},e)$ where $G$ and ${G_1}$ are two cyclic additive groups of prime order $q$, $g$ is a generator of $G$.
\item \textbf{Step 2:} The SO generates $({G_2},q,{g_2})$ where ${G_2}$ is a cyclic additive group of prime order $q$ and generator ${g_2}$ based on elliptic curves.
\item \textbf{Step 3:} The SO chooses a full-domain hash function ${H_1}:{\{ 0,1\} ^ * } \to G^2$ and a hash function ${H_2}$.
\item \textbf{Step 4:} The SO publishes public parameters $(q,g,{g_2},\\G,{G_1},{G_2},e,{H_1},{H_2})$.

\end{itemize}

\subsubsection{The first layer’s weights of the model}The SO trains the electricity theft detection model based on historical honest and malicious consumers' power consumption data, and then saves the weight of the first layer of the network, the weight $W = [w_1^{\rm T},w_2^{\rm T}, \cdots w_n^{\rm T}]$ can be represented as:

$$W = {\left[ {\begin{array}{*{20}{c}}
{{w_1}[1]}&{{w_2}[1]}& \cdots &{{w_n}[1]}\\
{{w_1}[2]}&{{w_2}[2]}& \cdots &{{w_n}[2]}\\
 \vdots & \vdots & \cdots & \vdots \\
{{w_1}[d]}&{{w_2}[d]}& \cdots &{{w_n}[d]}
\end{array}} \right]_{d \times n}}$$
where $d$ is the number of power reporting in the electricity theft detection period $T = \{ {t_1},{t_2}, \cdots {t_d}\}$ and $n$ is the number of neurons in the first layer of the neural network, $n$ should be fewer than the number of inputs $d$, because if $n \ge d$, the SO will calculate the consumers' fine-grained power consumption data, since $d$ unknowns in $d$ equations may be solved to obtain the data.

\subsubsection{Key Generation}SM generates the encryption keys and decryption keys. All SMs $\mathcal{U} = \{S{M_1},S{M_2},\cdots,S{M_m}\} $ cooperate using the secure aggregation algorithm to generate a decryption public key $DA$ for aggregating the power consumption readings of all SMs. Meanwhile, each SM generates electricity theft detection public keys $DW$.
%The ${\rm{S}}{{\rm{M}}_i}$ randomly select ${x_i} \in Z_q^*,(i = 1,2, \cdots ,n)$ as the private keys and $Pu{b_i} = {x_i} \cdot g$ as the public key and made public;

\begin{itemize}
\item Secret key generation: $S{M_i}$ selects a random number ${x_i} \in {Z_q}$ as the secret key for signing and key negotiation and selects ${s_i} \in Z_q^2$ as the secret key for encryption.
\item Generation of $DA$: Arbitrary SMs negotiate key masks among themselves and the MN performs secure aggregation to generate decryption public keys $DA$.

\textbf{Step 1:} Each ${SM_i}$ calculates and publishes the public key $s_i^{pk}={x_i}{g_2}$.\\
\textbf{Step 2:} Each ${SM_i}$ receives the public keys $s_o^{pk}$ of other SMs and then calculates ${sk_{i,o}} = {x_i}s_o^{pk}(i \in \mathcal{U};o \in \mathcal{U},o \ne i )$. Fig. 2 shows an example of four SMs performing key masks agreement and generating DA.\\
\textbf{Step 3:} Each ${SM_i}$ calculates ${y_i}$ and sends the results to MN for aggregation by Eq. (3).
\begin{equation} 
{y_i} = {s_i} + \sum\limits_{o \in \mathcal{U},i < o} {{sk_{i,o}}}  - \sum\limits_{o \in \mathcal{U},i > o} {{sk_{o,i}}} 
\end{equation} 
\textbf{Step 4:} MN aggregates its own $y$ and the results sent by other SMs, as shown in Eq. (4).

\begin{figure}[!ht]
\centering
\includegraphics[width=1.9in]{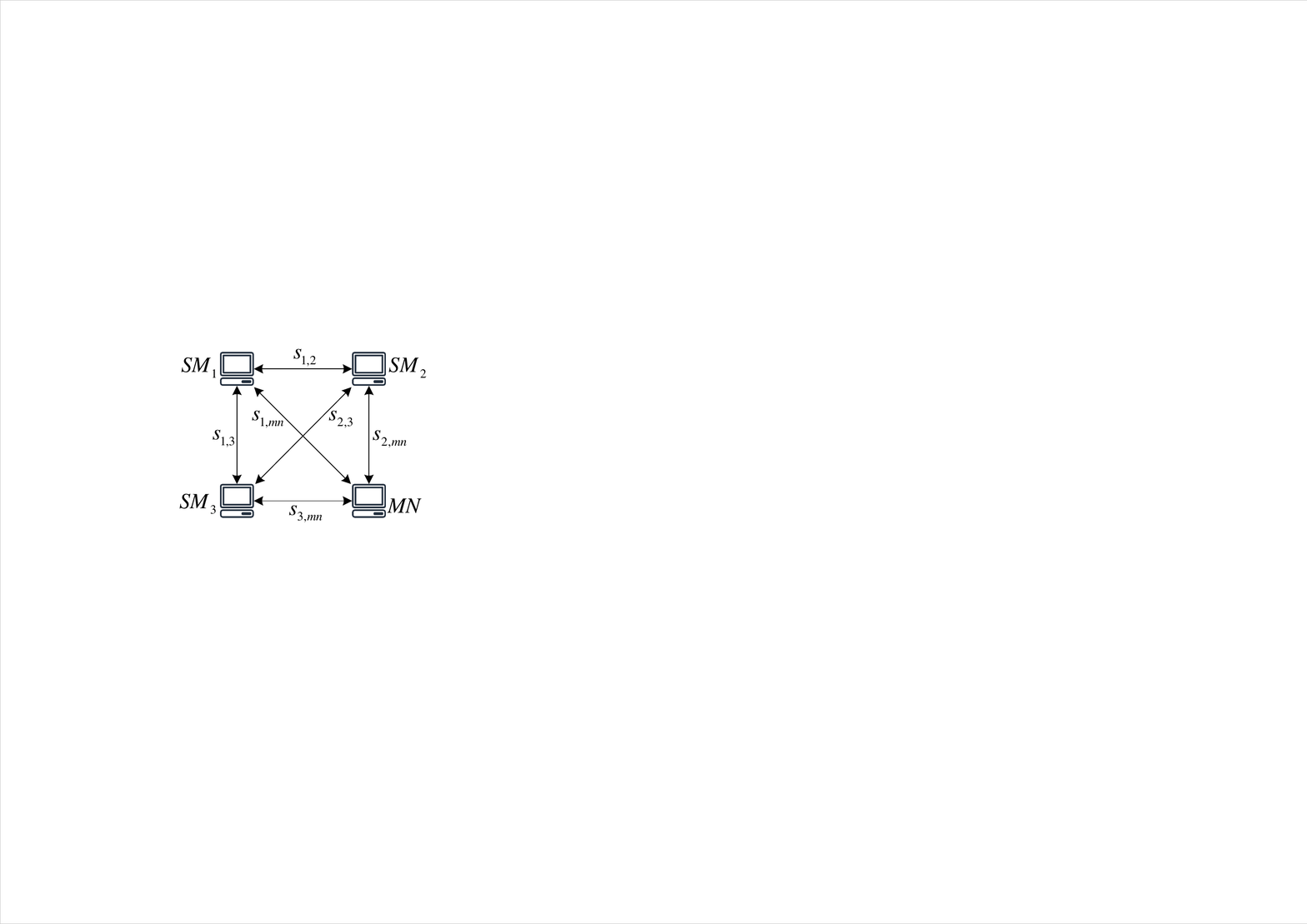}
\caption{Example of four SMs key masks agreement and generate DA.}
\end{figure}

\begin{equation} 
  \begin{aligned}
  DA &= \sum\limits_{i \in \mathcal{U}} {{y_i}}  \hfill \\
    &= \sum\limits_{i \in \mathcal{U}} {({s_i} + \sum\limits_{o \in \mathcal{U},i < o} {{sk_{i,o}}}  - \sum\limits_{o \in \mathcal{U},i > o} {{sk_{o,i}}} )}  \hfill \\
    &= \sum\limits_{i \in \mathcal{U}} {{s_i} \in Z_q^2}  \hfill \\ 
  \end{aligned} 
\end{equation} 

\item Generation of $DW$: SO publishes the weights of the first layer network of the electricity theft detection model to each ${SM_i}$, and each ${SM_i}$ generates decryption public keys to enable theft detection without obtaining the original power consumption data.\\
\textbf{Step 1:} Each ${SM_i}$ generates a timestamp $TS_t^i$ of the current detection time $T = \{ {T_1},{T_{2,}} \cdots {T_d}\}$ by Eq. (5).
\begin{equation} 
TS_t^i = {H_1}({T_t}) \in {G^2},t = \{ 1,2, \cdots ,d\}
\end{equation}
\textbf{Step 2:} Each ${SM_i}$ generates $D{W_{ci}},c = \{ 1,2, \cdots ,n\} $ decryption public keys by Eq. (6):
\begin{equation} 
D{W_{ci}} = \sum\limits_{t = 1}^d {{w_c}[t]} (s_i^ \top  \cdot TS_t^i) \in G
\end{equation}
\textbf{Step 3:} Each ${SM_i}$ generates decryption public keys by Eq. (7).
\begin{equation} 
D{W_i} = \{ D{W_{1i}},D{W_{2i}}, \cdots ,D{W_{ni}}\} 
\end{equation}

\end{itemize}

\subsection{Reporting Phase}
In the reporting phase, each ${SM_i}$ encrypts its power consumption readings and then performs signature operations.
\begin{itemize}
\item \textbf{Step 1:} For each electricity theft detection period $T$, each ${SM_i}$
encrypts its power consumption readings by Eq. (8).
\begin{equation} 
{\mathbb{C}_i}[t] = (s_i^ \top  \cdot TS_t^i) + {r_i}[t]g \in G
\end{equation}
\item \textbf{Step 2:} Each ${SM_i}$ computes the public key $P{K_i}= {x_i} \cdot {g_2}$ and then generates the BLS short signature by Eq. (9), $TS_t^i$ is the current timestamp to prevent replay attack.
\begin{equation} 
{\delta _i} = {x_i} \cdot {H_2}({\mathbb{C}_i}[t]||TS_t^i||D{W_i}||P{K_i})
\end{equation}
\item \textbf{Step 3:} Each ${SM_i}$ sends ${\mathbb{C}_i}[t]||TS_t^i||D{W_i}||{\delta _i}||P{K_i}$ to MN. The data within each ${SM_i}$ consists of basic storage information and primary transmission data, as shown in Fig. 3.
\end{itemize}

\begin{figure}[!t]
\centering
\includegraphics[width=2.7in]{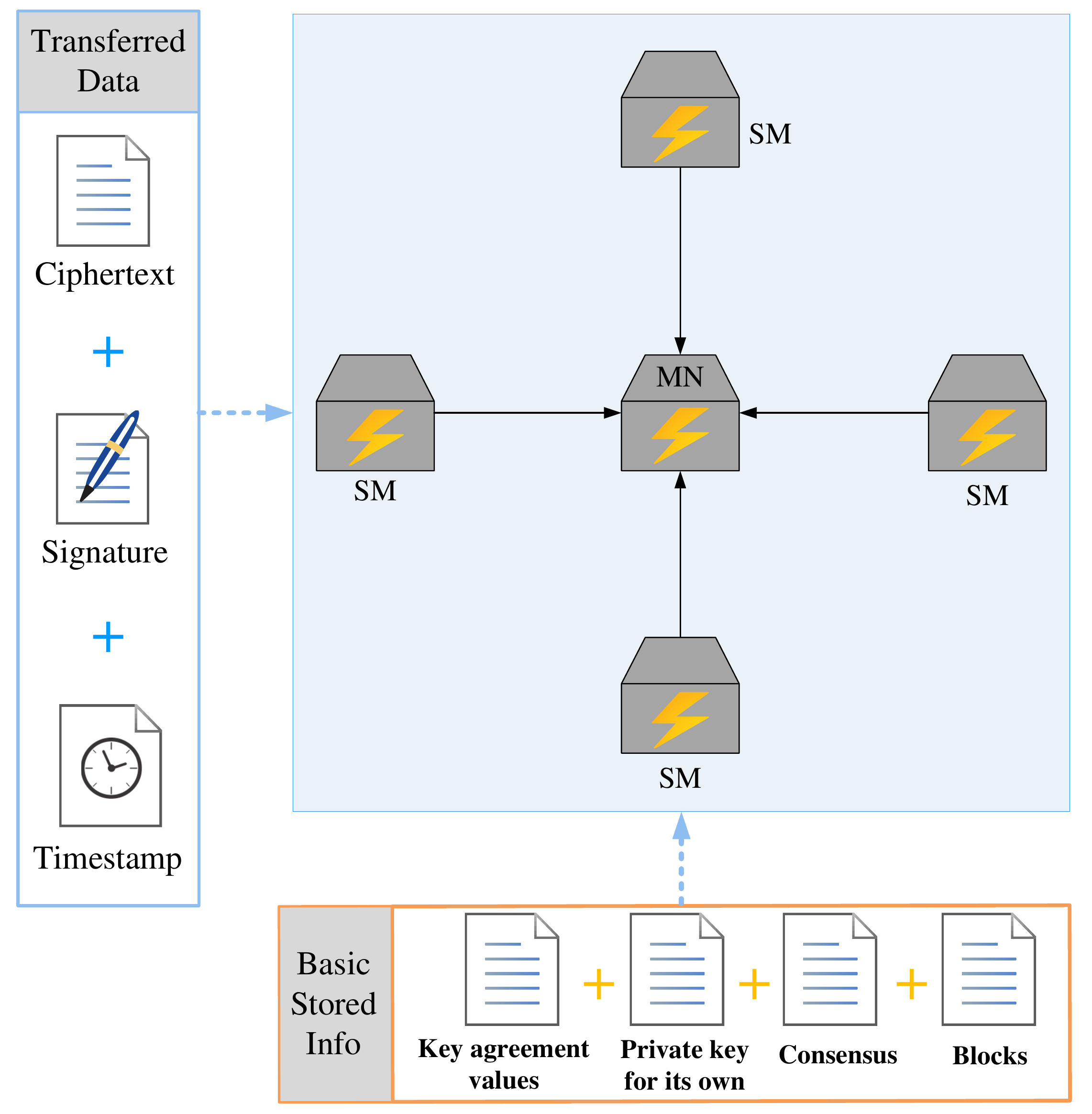}
\caption{The data within each smart meter node consists of basic stored information and primary transmitted data.}
\end{figure}

\subsection{Aggregating Phase}
Efficient message propagation methods are important building blocks for various networks\cite{zou2021jamming}, and in the proposed scheme, the SM sends messages directly to the MN, which is responsible for broadcasting and aggregating the total area power consumption. After receiving the data from the SMs, first, the $M{N_j},j \in |RA|$ in the residential area $R{A_j}$ verifies the signature and timestamp. After the verification is passed, $M{N_j}$ generates the Merkle tree and then creates the block through the Byzantine fault-tolerant consensus mechanism, the block head stores the timestamp, the hash of the previous block, and the Merkle tree root hash, and the block body stores the encrypted data and decryption Keys. After that, $M{N_j}$ aggregates the ciphertext and decrypts it to get the total power consumption of all SMs at the current time. Fig. 4 shows the blockchain structure of the proposed scheme. The detailed steps are as follows:
\begin{itemize}

\item \textbf{Step 1:} $M{N_j}$ verifies signature and timestamp. If Eq. (10) and $TS_t^i = TS_t^{MN}$ are valid, the verification passes and fails otherwise. To make verification more efficient, $M{N_j}$ can perform batch verification.
\begin{equation} 
e({\delta _i},{g_2}) = e({H_2}({\mathbb{C}_i}[t]||TS_t^i||D{W_i}||P{K_i}),P{K_i})
\end{equation}
\item \textbf{Step 2:} $M{N_j}$ performs the hash operation to generate the Merkle tree root hash value. Then $M{N_j}$ generates a new block $Block = (Data||H(Data||Timestamp)||Timest\\amp)$, and broadcasts the block to other SMs in the residential area $R{A_j}$. 
\item \textbf{Step 3:} After receiving the block, SMs verify the block's hash value, timestamp, and data, then send the result of the verification to other SMs to achieve mutual supervision among SMs. 
\item \textbf{Step 4:} SMs send their own check results to $M{N_j}$. $M{N_j}$ collects feedback from all SMs and checks them. If all SMs agree on the legitimacy and integrity of the block, $M{N_j}$ adds the block to the blockchain in chronological order and sends the block to other SMs. If there exists SM disagrees with the check result, $M{N_j}$ checks the feedback information and sends the block to this SM again for a second check. 
\item \textbf{Step 5:} $M{N_j}$ aggregates the encrypted data of all SMs and decrypts it to get the total power consumption of the area at the current time by Eq. (11). 

\end{itemize}

Since  $\sum\limits_{i \in \mathcal{U}} {r_i}[t]g$ is not a very large value, there are many ways to calculate the aggregated value, such as Shank’s baby-step giant-step algorithm\cite{shoup1997lower}.

\begin{equation} 
\begin{gathered}
\sum\limits_{i \in \mathcal{U}} {{\mathbb{C}_i}[t]}  - D{A^ \top }T{S_t} \hfill \\
 = \sum\limits_{i \in \mathcal{U}} {((s_i^ \top  \cdot T{S_t}) + {r_i}[t]g)}  - {(\sum\limits_{i \in \mathcal{U}} {{s_i}} )^ \top }T{S_t} \hfill \\[1mm]
 = \sum\limits_{i \in \mathcal{U}} {{r_i}[t]g \in G}  \hfill \\[1mm]
\end{gathered} 
\end{equation}

\begin{figure}[!t]
\centering
\includegraphics[width=3.6in]{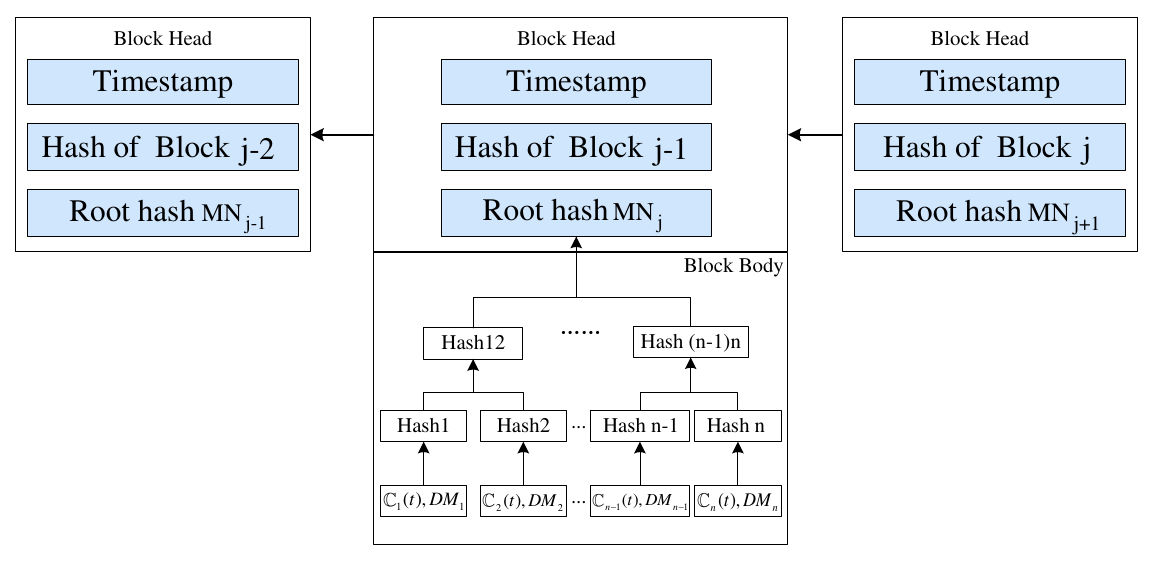}
\caption{The blockchain structure of the proposed scheme.}
\end{figure}

\subsection{Judgement Phase}
To achieve efficient detection, our solution will perform electricity theft detection after discriminating whether there is electricity theft in residential areas.

For each residential area, the transformer meter measures the total amount of electricity supplied to that residential area during the electricity theft detection period, ${E_{DTM}}(t)$. Meanwhile, the MN aggregates the readings uploaded by all SMs in the residential area, $\sum\nolimits_i {{E_{S{M_i}}}(t)} $, the SO determines whether there is electricity theft by Eq. (12):

\begin{equation} 
\label{deqn_ex2}
{E_{DTM}}(t) > \sum\nolimits_i {{E_{S{M_i}}}(t)}  + {E_{TL}}(t) + \varepsilon 
\end{equation}
where ${E_{TL}}(t)$ is the technical loss (TL) in transmission lines within the residential area and $\varepsilon $ is the calculation error for TL. The SO can use historical data to analyze the technical loss, while many methods exist \cite{jokar2015electricity} to calculate the technical loss. If Eq. (12) is valid, SO considers that there is electricity theft in the current area. Afterward SO reads the power consumption data uploaded by each ${SM_i}$ in the blockchain for electricity theft detection.

\subsection{Electricity Theft Detection Phase}

In this sub-section, a privacy-preserving electricity theft detection model is presented in the proposed scheme, and then we explain the experimental settings, including computing platforms, dataset, and data pre-processing.

\subsubsection{\textbf{Privacy-preserving Electricity Theft Detection Model}} As shown in Fig. 5, our model is composed of the fully connected layer and long short-term memory networks. The core operation of the fully connected layer is the multiplication of a matrix and a vector, which can be expressed as $\bm{xW}$. More detailed representations are:
\begin{equation} 
\begin{gathered}
 \relax [{x_1},{x_2}, \cdots {x_d}] \times \left[ {\begin{array}{*{20}{c}}
  {{w_1}[1]}&{{w_2}[1]}& \cdots &{{w_c}[1]} \\ 
  {{w_1}[2]}&{{w_2}[2]}& \cdots &{{w_c}[2]} \\ 
   \vdots & \vdots & \cdots & \vdots  \\ 
  {{w_1}[d]}&{{w_2}[d]}& \cdots &{{w_c}[d]} 
\end{array}} \right] \hfill \\
   = \bm{[x \cdot w_1^ \top ,x \cdot w_2^ \top , \cdots ,x \cdot w_c^ \top ]} \hfill \\ 
\end{gathered}  
\end{equation}
where $\boldsymbol{x}$ is the input vector, $\boldsymbol{W}$ is the weight matrix, and then $\bm{b}$ is the bias vector is added.  This operation can be seen as an inner product of the input vector $\bm{x}$ and each column of the weight matrix $\bm{W}$. It can also be viewed as a group of $n$ $d$-equations, where the input vector $\bm{x}$ are the unknowns and the weight matrix $\bm{W}$ are the coefficients, and since $n$ is less than $d$, the input vector $\bm{x}$ cannot be solved.

Therefore, in order to perform electricity theft detection in the ciphertext state of the consumers' power consumption data, the result of the inner product of consumers' power consumption data ${r_i} = [{r_i}[1],{r_i}[2], \cdots ,{r_i}[d]]$ and each column of the weight matrix is obtained by Eq. (14). 
\begin{equation} 
\begin{gathered}
  \sum\limits_{t = 1}^d {{w_c}[t]}  \times {\mathbb{C}_i}[t] - D{W_{ci}} \hfill \\
   = \sum\limits_{t = 1}^d {{w_c}[t]} ((s_i^ \top  \cdot T{S_t}) + {r_i}[t]g) - \sum\limits_{t = 1}^d {{w_c}[t]} (s_i^ \top  \cdot T{S_t}) \hfill \\
   = (\sum\limits_{t = 1}^d {{r_i}[t]{w_c}[t]})g \hfill \\ 
\end{gathered} 
\end{equation}
\begin{figure}[!t]
\centering
\includegraphics[width=2.5in]{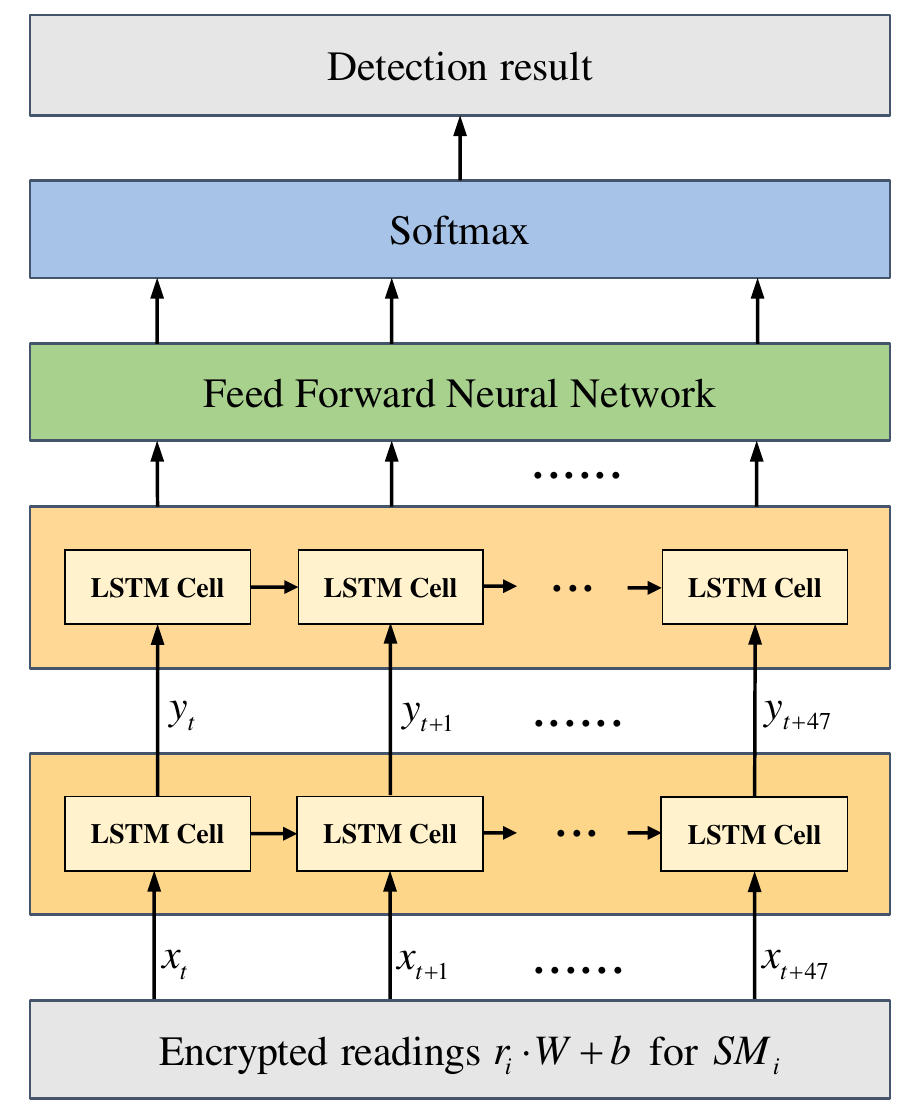}
\caption{The LSTM-based privacy-preserving electricity theft detection framework.}
\end{figure}

The output of the fully connected layer is obtained by calculating the inner product of each column of the weight matrix with the consumer power consumption data, and then adding the bias vector $\boldsymbol{b}$ as follows:

$
[(\boldsymbol{r_i} \cdot \boldsymbol{w_1^{\rm T}}) + \boldsymbol{b_1},(\boldsymbol{r_i} \cdot \boldsymbol{w_2^{\rm T}}) + \boldsymbol{b_2}, \cdots ,(\boldsymbol{r_i} \cdot \boldsymbol{w_c^{\rm T}}) + \boldsymbol{b_c}]
$

After SO gets the output result of the fully connected layer, it still cannot solve the original consumers' power consumption data, and the consumers' power consumption data is input to the next layer of the network in the encrypted state, finally, the detection result is inferred after layer-by-layer computation.

The detection model uses categorical cross-entropy as the loss function. In the model training phase, we use the RMSprop optimizer to train the model for 30 epochs, 512 batch sizes, and 0.001 learning rate. To prevent overfitting, we use the kernel $\ell 2 - $regularizer in the LSTM layer, and at the same time the callback function ReduceLROnPlateau in the Keras framework\cite{keras2015theano} is used to dynamically reduce the learning rate, and the callback function EarlyStopping is used to obtain the optimal model.
The parameters of our model structure are summarized in Table III, where AF stands for activation function.

\begin{table}[!ht]
\renewcommand\arraystretch{1.3}
\caption{Parameters of model structure\label{tab:table3}}
\centering
\begin{tabular}{|c|c|c|c|}
\hline
Layer(type) & No. of neurons & No. of parameters & AF\\
\hline
dense(Dense) & 10 & 20 & tanh\\
\hline
lstm(LSTM) & 300 & 373200 & tanh,sigmoid\\
\hline
lstm-1(LSTM) & 300 & 721200 & tanh,sigmoid\\
\hline
dense-1(Dense) & 2 & 602 & softmax\\
\hline
\end{tabular}
\end{table}
\subsubsection{\textbf{Computing Platforms}}In our experiments, we build a Tensorflow virtual environment on a server with unbutu 18.04.6 LTS system and NVIDIA Tesla T4 GPU as well as use the Keras framework to train and evaluate the model.

\subsubsection{\textbf{Dataset}}We use the dataset from the Irish Smart Energy Trials \cite{ref30}, which contains the power consumption data of more than 1000 consumers in 535 days from 2009 to 2010, and fine-grained power consumption data is reported by each SM every 30 minutes.

\subsubsection{\textbf{Data Pre-processing}}We select the smart meter data of 200 consumers from the dataset and create one record of the consumer's power consumption data (48 readings) for one day, with a total of 107,200 records. 

Since all the data in the dataset are from honest consumers' data, we use the electricity theft attack proposed by \cite{jokar2015electricity} to generate malicious consumers' data.  Based on the dataset of benign samples, for each sample $X = \{ {x_t}|1 \le t \le 48\} $,  we perform the following operations to generate six malicious types of data:

\begin{enumerate}
\item ${f_1}({x_t}) = \alpha {x_t}$, $\alpha  = random(0.1,0.8)$;
\item ${f_2}({x_t}) = {\beta _t}{x_t}$, ${\beta _t} = random(0.1,0.8)$;
\item ${f_3}({x_t}) = mean(X)$;
\item ${f_4}({x_t}) = {\beta _t}mean(X)$, ${\beta _t} = random(0.1,0.8)$;
\item ${f_5}({x_t}) = {x_{48 - t}}$;
\item ${f_6}({x_t}) = {\gamma _t}{x_t}$. 

${\gamma _t} = \left\{ {\begin{array}{*{20}{c}}
0&{ts < t < te}\\
1&{else}
\end{array}} \right.$
$\begin{array}{l}
ts = random(0,42)\\
te - ts = random(6,48)
\end{array}$

\end{enumerate}

Electricity theft attacks generated 643,200 malicious data records. Since the data of honest data records are only 107200, this leads to the problem of unbalanced sample categories of data. Therefore, we apply the adaptive synthetic sampling method (ADASYN) \cite{ref31} to balance the size of honest and malicious classes. We randomly divide the balanced dataset into a training dataset (80\%) and a testing dataset (20\%) to perform the training of the model.

\begin{figure*}[!t]
\centering
\subfloat[]{\includegraphics[width=2in]{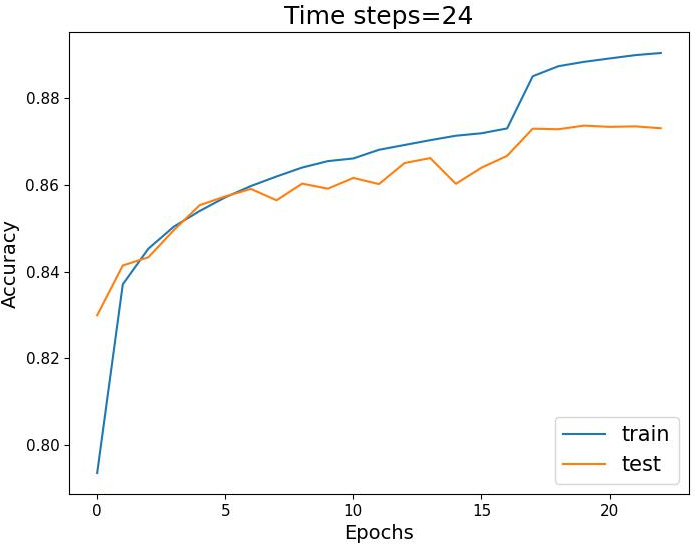}%
\label{fig_first_case}}
\hfil
\subfloat[]{\includegraphics[width=2in]{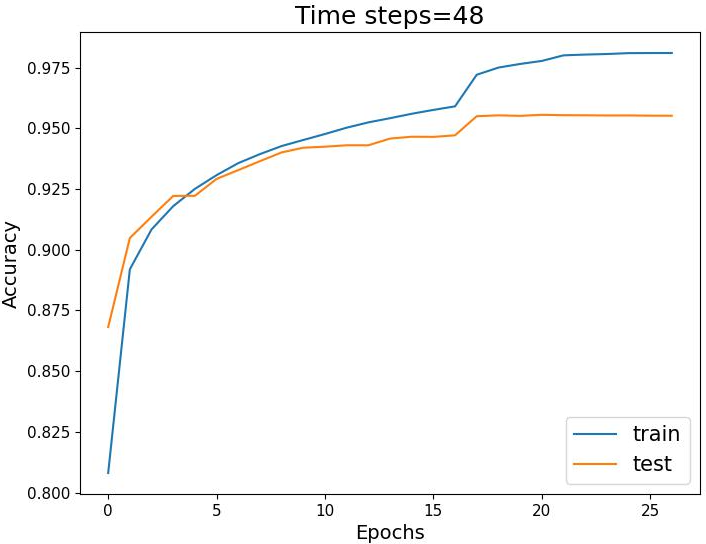}%
\label{fig_second_case}}
\hfil
\subfloat[]{\includegraphics[width=2in]{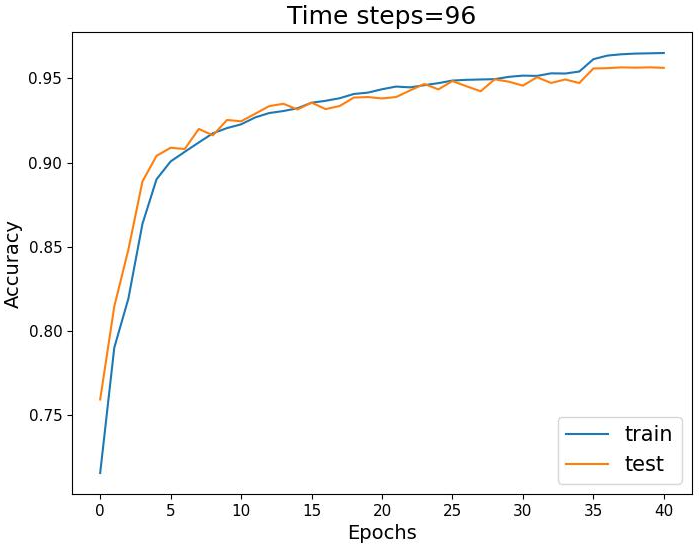}%
\label{fig_third_case}}
\caption{Parameter study of time steps $t$.}

\end{figure*}

\begin{figure*}[!t]
\centering
\subfloat[]{\includegraphics[width=2in]{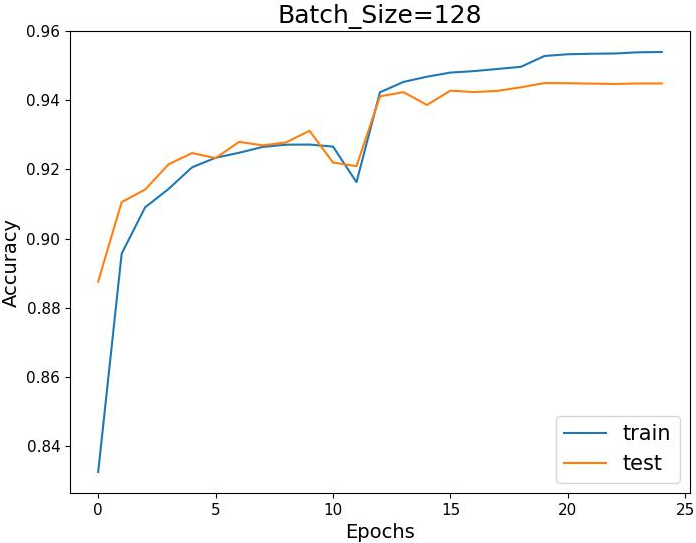}%
\label{fig_first_case1}}
\hfil
\subfloat[]{\includegraphics[width=2in]{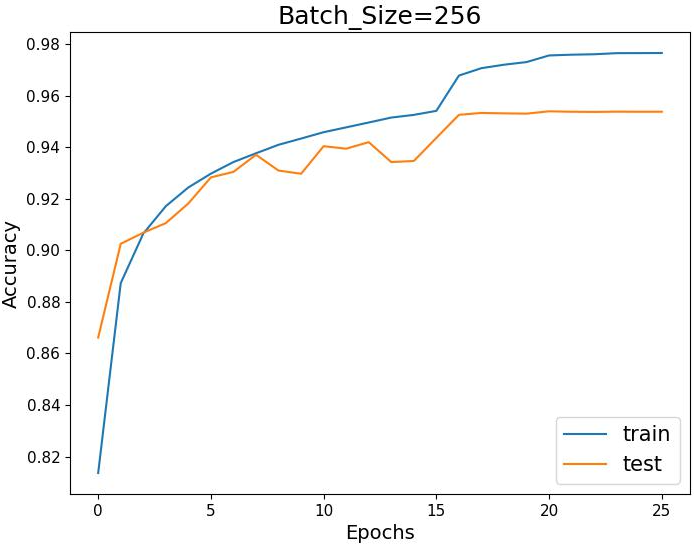}%
\label{fig_second_case1}}
\hfil
\subfloat[]{\includegraphics[width=2in]{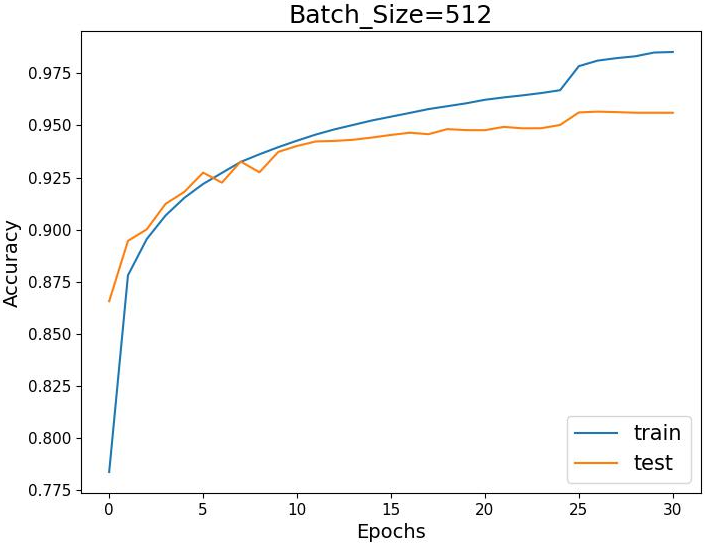}%
\label{fig_third_case1}}
\caption{Parameter study of batch size $\beta$.}

\end{figure*}

\section{Performance Evaluation}
In this section, at first, our method is compared with other methods that deal with time series to demonstrate the better performance of our method. Then, we study the parameters of our model. Finally, we evaluate the performance of our electricity theft detection model in our test set. Meanwhile, we compare the computation and communication of the model with other schemes.

\subsection{Method Comparison}
To demonstrate the better performance of our model, the experimental comparison with other methods was performed on a test dataset. Concretely, Deng et al. proposed a tree-ensemble method, referred to as time series forest (TSF), for time series classification \cite{deng2013time}. Middlehurst et al. proposed an improved hierarchical vote collective of transformation-based ensembles (HIVE-COTE) for time series classification \cite{middlehurst2021hive}. Dempster et al. proposed a simple linear classifier based on the random convolution kernels (ROCKET) \cite{dempster2020rocket}. Meanwhile, in \cite{nabil2019ppetd}, the authors proposed to use the one-dimensional convolutional neural network (CNN) for electricity theft detection. Table IV gives the experimental results for each method using the same training data set and testing data set, and we see that the LSTM model gets the highest accuracy score of $95.56\%$. 

\begin{table}[!t]
\renewcommand\arraystretch{1.3}
\caption{Algorithm Accuracy Scores\label{tab:table4}}
\centering
\begin{tabular}{|c|c|}
\hline
\rowcolor{gray!45} 
\textbf{Algorithm} & \textbf{Accuracy Score ($\%$)}\\
\hline
The LSTM model & \textbf{95.56}\\
\hline
1-D CNN model \cite{nabil2019ppetd} & 93.20\\
\hline
Time Series Forest \cite{deng2013time} & 86.36\\
\hline
The improved HIVE-COTE \cite{middlehurst2021hive} & 90.91\\
\hline
ROCKET \cite{dempster2020rocket}  & 78.76\\
\hline
\end{tabular}
\end{table}

\subsection{Parameter Study}
Various hyper-parameters of the model have an impact on the performance of the model. For our model, what is more important is the time step, which is the number of power readings input in the model. In our model, the time step is the same as the theft detection period. For the theft detection model, increasing the detection period means that the communication overhead of the model will increase, so a reasonable theft detection period must be determined. Therefore, we deeply analyze the impact of these parameters on the performance of our model.

\subsubsection{\textbf{Effect of time steps $t$}}Fig. 6 shows the accuracy of the validation set with varying epochs when the time steps are different. We can find out that different time steps affect the accuracy of the model as well as the training time, while the longer the time steps, the longer the theft detection period will be, which will lead to a rise in the overall model in terms of communication overhead. Although the difference in accuracy between time steps 96 and 48 is not significant, the training time is shorter and communication is less when the time steps are 48.

\subsubsection{\textbf{Effect of learning rate $\ell $}}In the model training progress, we use the RMSprop optimizer with a default learning rate $\ell =0.001 $. To find the optimal model, we use the callback function ReduceLROnPlateau in the Keras framework, which serves to reduce the learning rate when learning stagnates. As shown in Fig. 6, there is some improvement in accuracy after reducing the learning rate.

\subsubsection{{\bf{Effect of batch size $\beta$:}}}Fig. 7 shows the performance of our model by setting the batch size as 512 which gets the highest accuracy with 95.60 $\%$ while needing more epochs to optimize. The experimental results show that a smaller batch size can speed up the optimization within the same epochs, which suggests that setting the bath size between 256 and 512 is more acceptable.

\subsubsection{\textbf{Effect of neurons $\eta $}}Fig. 8 shows that the highest accuracy is achieved when the number of neurons in the LSTM layer is 300-360. A more number of model neurons represents a slower model inference, so the neurons of our model are set to 300.

\begin{figure}[!t]
\centering
\includegraphics[width=2.5in]{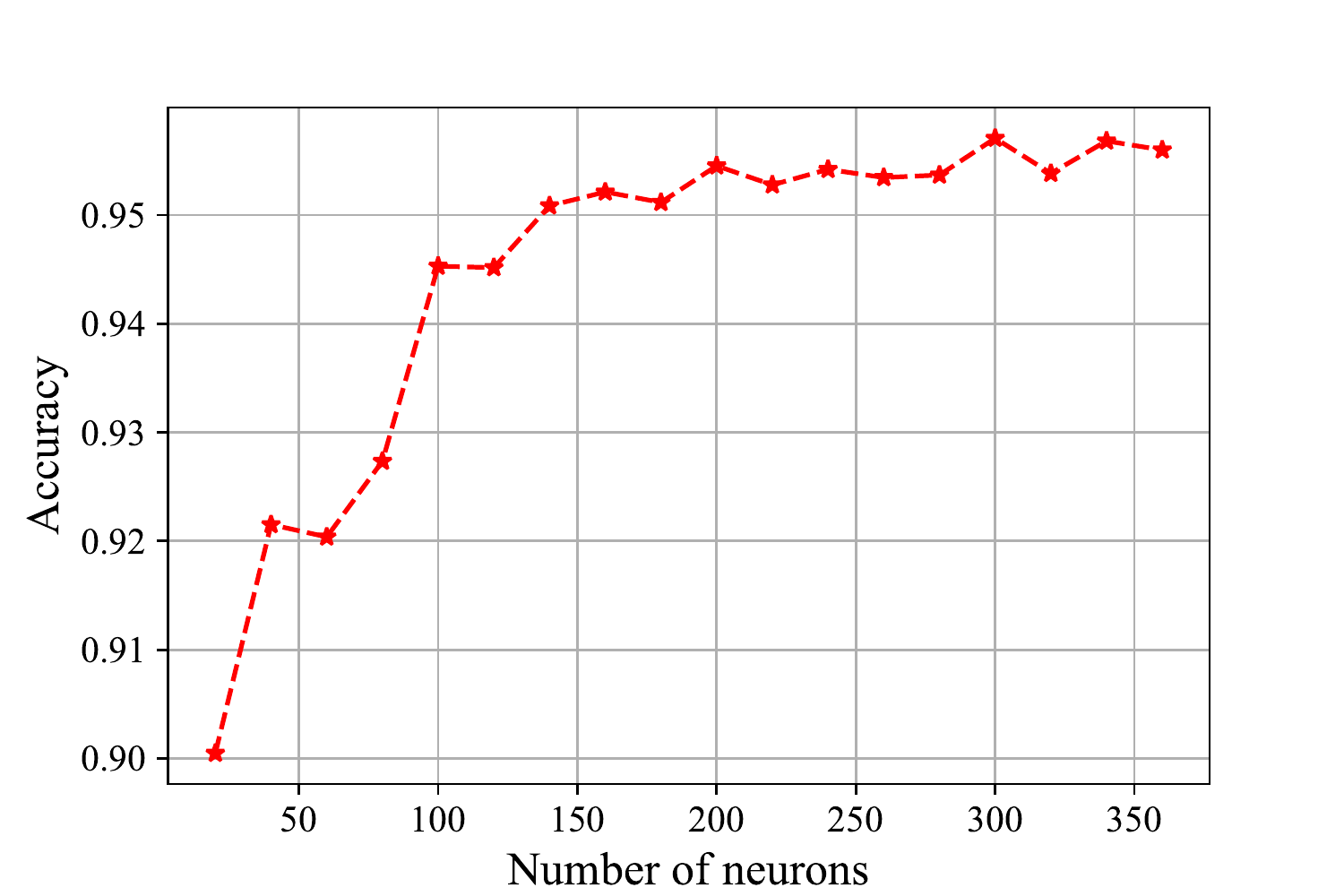}
\caption{Parameter study of neurons $\eta $.}
\end{figure}

\begin{figure}[!t]
\centering
\includegraphics[width=2.5in]{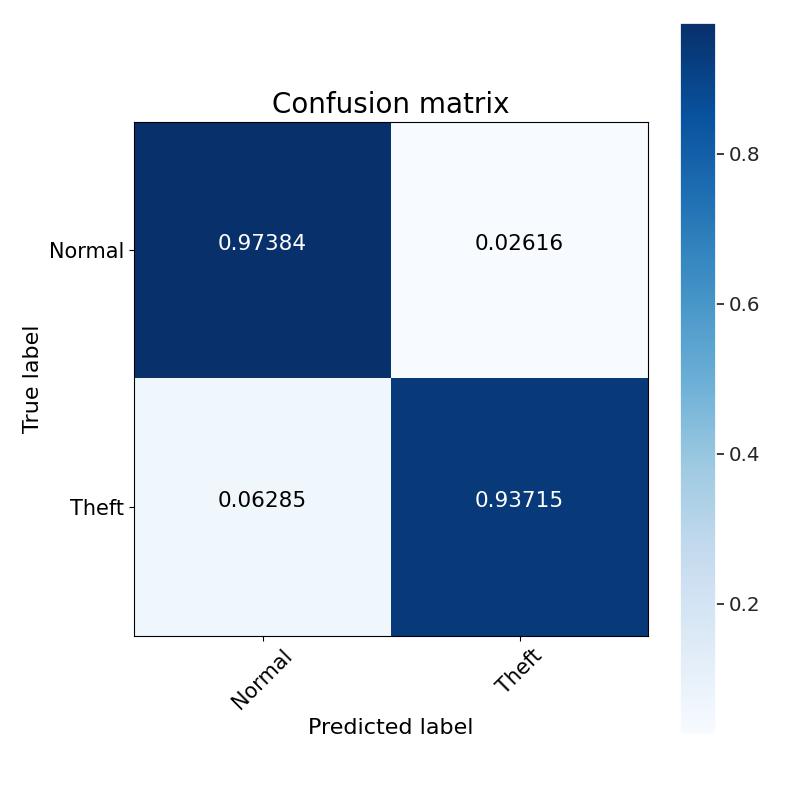}
\caption{Confusion matrix of our model.}
\end{figure}

\begin{table*}[!t]
\renewcommand\arraystretch{1.5}
\caption{The performance of our model and other schemes\label{tab:table5}}
\centering
\begin{tabular}{|c|c|c|c|c|c|c|}
\hline 
\rowcolor{gray!45} 
\textbf{Model} & \textbf{DR($\%$)} & \textbf{FA($\%$)} & \textbf{HD($\%$)}  & \textbf{Accuracy($\%$)} & \textbf{Model detection overhead} & \textbf{Communication overhead}\\
\hline
\textbf{Our model} & 93.72 & \textbf{2.62} & \textbf{91.10} & \textbf{95.56} & \textbf{56.03 ms} & 600 Bytes\\
\hline
\textbf{ETDFE \cite{ibrahem2020efficient}}& 92.56 & 5.84 & 86.72 & 93.36 & 1.94 seconds & 40 Bytes\\
\hline
\textbf{PPETD MD1 \cite{nabil2019ppetd}} & 91.50 & 7.40 & 84.10 & 91.80 & 48 minutes & 1900 MB\\
\hline
\textbf{PPETD MD2 \cite{nabil2019ppetd}} & 90.00 & 8.79 & 81.2 & 90.20  & 39 minutes & 1675 MB\\
\hline
\textbf{PPETD MD3 \cite{nabil2019ppetd}} & 88.60 & 3.90 &84.60 & 90.30  & 35 minutes & 1375 MB\\
\hline
\textbf{Jokar et al\cite{jokar2015electricity}} & \textbf{94.00} & 11.0& 83.0 & - & - & - \\
\hline
\end{tabular}
\end{table*}

\subsection{Performance of electricity theft detection model}
\subsubsection{\textbf{Performance Metrics}}To evaluate our electricity theft detection model, we conduct the experiments by considering four performance metrics: accuracy, the detection rate $(DR)$, and the false acceptance rate $(FA)$ as well as the highest difference $(HD)$. Accuracy measures the percentage of correct classifications in the testing dataset. The detection rate measures the percentage of detected malicious consumers in the total malicious consumers. The false acceptance rate measures the percentage of honest consumers who are mistakenly detected as malicious consumers. When $DR$, accuracy, and $HD$ are high and $FA$ is low, the model performance is better.

$$DR = \frac{{TP}}{{TP + FN}}, FA = \frac{{FP}}{{TN + FP}},HD = DR - FA,$$
where $TP$, $FP$ and $TN$ stands for true positive, false positive
and true negative, respectively.

$$accuracy = \frac{1}{s}\sum\limits_{i = 1}^s {\phi (f({x_i}),{y_i})},$$

$$\phi (f({x_i}),{y_i}) = \left\{ {\begin{array}{*{20}{c}}
1&{f({x_i}) = {y_i}}\\
0&{f({x_i}) \ne {y_i}}
\end{array}} \right.,$$
where $s$ is the total number of samples in the  testing dataset, $y_i$ is the
label for the $i-$th consumer, $f({x_i})$ is is the inference result of the model.

\subsubsection{\textbf{Performance Comparison}}We obtain the confusion matrix of our model by using the Scikit-learn python library. As shown in Fig. 9, in the confusion matrix of our model, the proportion of consumers who are predicted to be electricity theft consumers among those who are electricity theft consumers is the $DR$, the proportion of consumers predicted to be electricity theft as a percentage of those who are truly normal consumers is the $FA$.

Table V shows the evaluation results of our model and the existing models with privacy preservation. The proposed scheme is better in terms of $FA$, accuracy, and $HD$ among the schemes considering privacy protection. Our privacy detection model has higher accuracy and $HD$, 95.56$\%$, and 91.10$\%$, respectively. At the same time, the $FA$ in our model is 2.62$\%$, which is lower than other schemes. From the evaluation results, we can demonstrate that the proposed scheme has a better performance. Moreover, the performance of our model is not decreased by the use of encryption compared to \cite{nabil2019ppetd} because we use the inner product operation of the parameters of the first layer of the model with the consumers' power consumption data and the operation result in the same output as the direct input to the model.

\subsection{Computation and Communication Overhead}
To evaluate the proposed scheme in a more realistic environment, we used the Python "Charm" crypto-graphic library \cite{akinyele2013charm} on a Raspberry Pi Zero W device with a 1.0 GHz single-core CPU and 512 MB of RAM. The elliptic curve of size 160 bits (MNT159 curve) was also used.
\subsubsection{\textbf{Communication overhead}}In our model, the main communication overhead comes from the SMs transferring ${\mathbb{C}_i}||TS_t^i||D{W_i}||{\delta _i}||P{K_i}$ to the MN. We use an elliptic curve with 160-bit security level. From Eq. (5) to Eq. (9), it can see that the ciphertext, signature, and public key $PK$ size are all 40 bytes, the $D{W_i} = \{ D{W_{1i}},D{W_{2i}}, \cdots ,D{W_{10i}}\}$  size is 400 bytes, and the timestamp size is 80 bytes, so it takes 600 bytes for the ${SM_i}$ to report one reading. PPETD \cite{nabil2019ppetd} uses secure multiplication, $sigmoid( \cdot )$ security evaluation, and garbled circuits to protect the privacy evaluation of the CNN model, which results in a high communication overhead of about 1900 MB per SM. Yao et al.'s scheme \cite{yao2019energy} requires sending a ciphertext, signature, and timestamp to two institutions to complete the aggregation and detection, and we assume that it generates 2048 bits of ciphertext, 40 bytes of signature, and 40 bytes of timestamp, the total size required is 672 bytes. Richardson et al.\cite{richardson2016privacy} and Ibrahem et al.'s scheme \cite{ibrahem2020efficient} only sends 40 bytes, and 256 bytes of ciphertext, respectively. Meanwhile, Fig. 10 gives a comparison of the communication overhead with other schemes. It can be seen that the proposed scheme achieves more security within an acceptable range of communication overhead.

\begin{figure}[!t]
\centering
\includegraphics[width=3.2in]{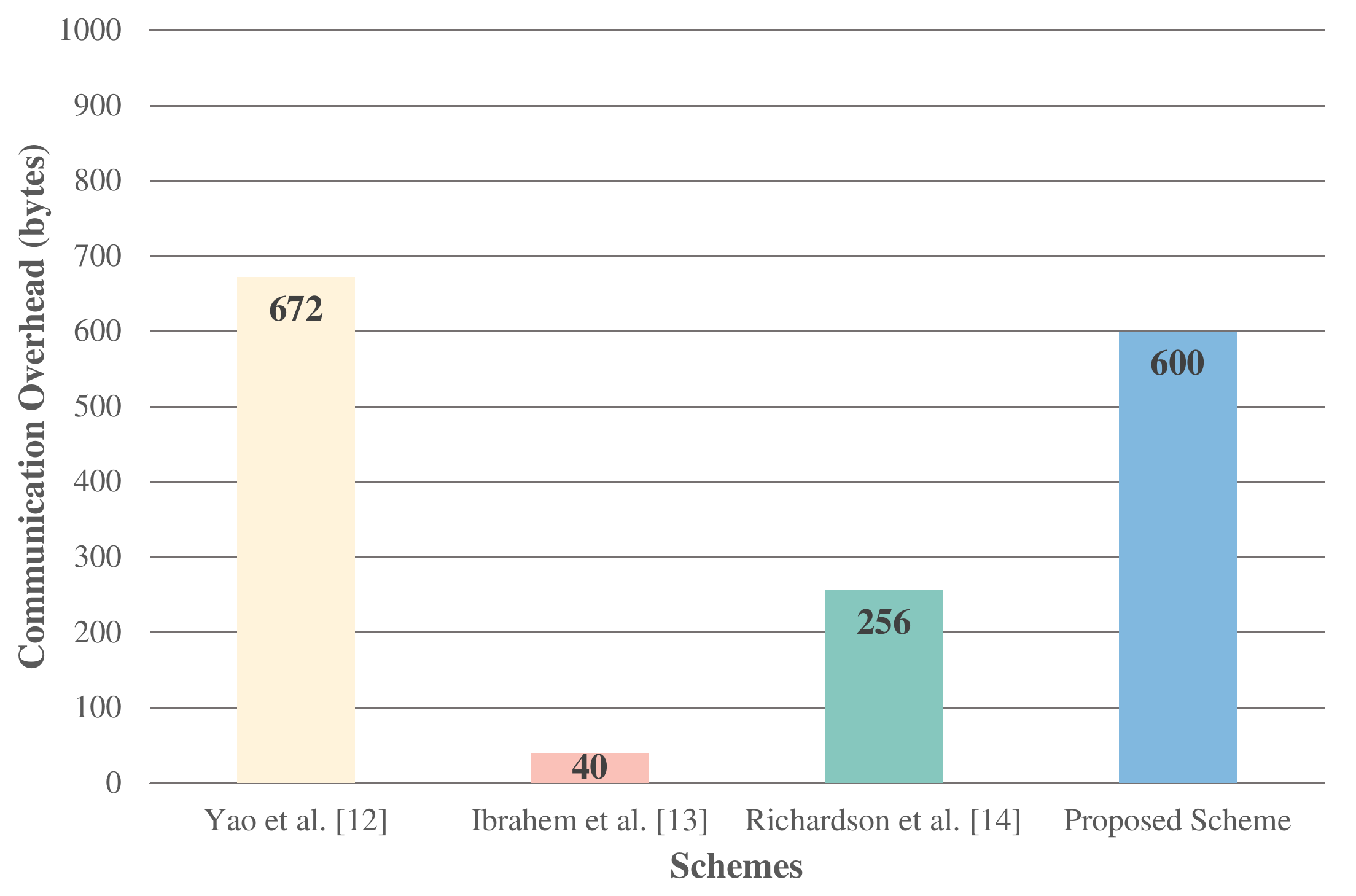}
\caption{Comparison of the communication overhead with other schemes.}
\end{figure}

\subsubsection{\textbf{Computation overhead}}In the proposed scheme, the computations mainly include three phases: reporting phase, aggregating phase, and electricity theft detection phase. In the reporting phase, the main computation overhead comes from the encryption, signature, decryption keys generation, and timestamp generation operations of the SM, therefore, the total time cost of the reporting phase is 59.488 ms. In the aggregation phase, MN achieves aggregating readings, decrypting, and verifying signatures, the total time cost is 129.635 ms. In the detection phase, the computation cost of decrypting to obtain $\bm{rW}$ is 49.63 ms. The computation costs of required functions are listed in Table VI. Experimental results show that this is feasible in a real-world environment.

In model inference speed, the total evaluation time of ETDFE for a 15-layer FFN model with 3,391,634 parameters is about 1.94 seconds and PPETD MD1 takes 48 minutes to evaluate the model, our model has only 1,095,022 parameters and its evaluation time is only 56.03 ms. In addition, the proposed scheme is more efficient compared to other schemes because it performs electricity theft detection after identifying the suspected theft area.

\begin{table}[!t]
\caption{Average computation cost of basic functions\label{tab:table6}}
\renewcommand\arraystretch{1.3}
\centering
\begin{tabular}{ccc}
\hline
Notations & Description & Time (ms)\\
\hline
${T_C}$ & Time cost of encryption & 0.096\\
${T_{agg}}$ & Time cost of aggregating 200 readings & 2.21\\
${T_{decAgg}}$ & Time cost of decrypting aggregated readings & 0.135\\
${T_{DM}}$ & Time cost of public key generation $DW$ & 45.36\\
${T_{decDW}}$ & Time cost of decrypting to obtain $\bm{rW}$ & 49.63\\
$T{S_t}$ & Time cost of generating timestamp & 0.852\\
${T_{sig}}$ & Time cost of signature operation & 13.18\\
${T_{versig}}$ & Time cost of the verify signature operation & 127.29\\
${T_m}$ & Time cost of model detection & 56.03\\
\hline
\end{tabular}
\end{table}

\subsection{Blockchain simulations}
The block time is a measure of the time it takes for the miners or validators in the network to verify the transactions within a block and generate a new block in that blockchain. Very short block times may lead to abnormal behavior, because nodes may not have enough time to send transactions, and synchronize their transaction pool or blockchain. Very long block time wastes arithmetic power and reduces the security of the system. Therefore, an appropriate block time is important. As shown in Fig. 11, average block time is simulated in the blockchain simulation system\cite{stoykov2017vibes} for the number of SMs in the detection region from 50 to 300. The block time should be as much as possible less than the period of the SM reporting power consumption readings, and the SO can select the number of SMs in the area based on the reporting period.

\begin{figure}[!t]
\centering
\includegraphics[width=3.1in]{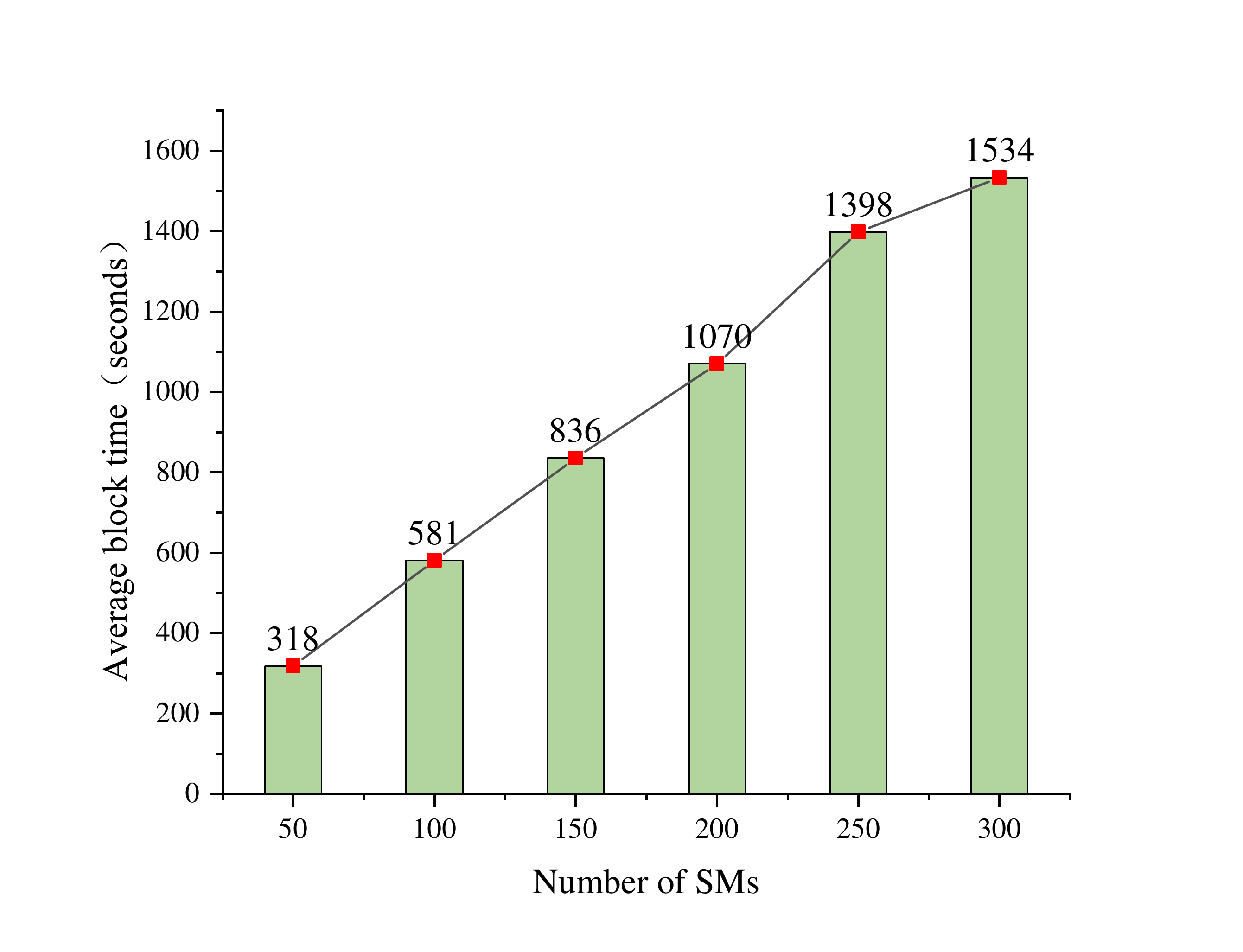}
\caption{Average block time for the number of SMs from 50 to 300.}
\end{figure}

\section{System Analysis}
In this section, we aim to demonstrate that the proposed scheme can achieve the following security and privacy guarantee, while it can resist the attacks in Section IV-B. In addition, to prove that the proposed scheme is more secure than the existing schemes, we perform a comparison of system characteristics.

\subsection{Security analysis}

\textbf{Scenario 1:} The privacy of consumers' power consumption cannot be inferred by any attacker.

\textbf{Proof:} The consumers' fine-grained data ${r_i}[t]$ is encrypted and sent to the MN. The conﬁdentiality of ${r_i}[t]$ is achieved by an elliptic curve over finite fields. Speciﬁcally, to analyze the consumers' private information, the attacker needs to crack the consumer's continuous long-term encrypted data $[ \cdots ,\mathbb{C}{_i}[t - 1],\mathbb{C}{_i}[t],\mathbb{C}{_i}[t + 1], \cdots ]$, but the attacker can only get the public parameters, which is infeasible in cracking the computation. In the electricity theft detection stage, the input encrypted power consumption data $\mathbb{C}{_i}[t],t = [1,2, \cdots ,d]$ is decrypted to get the output result $[(\boldsymbol{r_i} \cdot \boldsymbol{w_1^{\rm T}}) + \boldsymbol{b_1},(\boldsymbol{r_i} \cdot \boldsymbol{w_2^{\rm T}}) + \boldsymbol{b_2}, \cdots ,(\boldsymbol{r_i} \cdot \boldsymbol{w_n^{\rm T}}) + \boldsymbol{b_n}]$ of the first layer of neural network. Since $n$ is less than $d$, $n$  $d$-element equations cannot be solved, therefore the SO cannot obtain the original consumer's power consumption data $[{r_i}[1],{r_i}[2], \cdots ,{r_i}[d]]$, while still complete the electricity theft detection. Therefore, the proposed scheme preserves the privacy of consumers.

\textbf{Scenario 2:} Consumers' fine-grained power consumption data cannot be falsified and forged during transmission and storage, etc.

\textbf{Proof:} The messages ${\mathbb{C}{_i}[t]}||TS_t^i||D{W_i}||P{K_i}$ sent by each $S{M_i}$ in the scheme are BLS signed as ${\delta _i} = {x_i} \cdot {H_2}(\mathbb{C}{_i}[t]||TS_t^i||D{W_i}||P{K_i})$ to ensure the integrity of the data and prevent falsification. After accepting the message, the MN creates a block after establishing the Merkle tree, and each $S{M_i}$ can access the block to verify whether its data has been falsified. Meanwhile, since all data transfers in the blockchain have timestamps and cannot be changed when added to the blockchain, so the proposed scheme can resist data falsification and forgery.

\textbf{Scenario 3:} The proposed scheme does not require a third party and also can resist collusion attacks by smart grid entities.

\textbf{Proof:} In the proposed scheme, the whole process does not require the participation of a third party, which makes the scheme more reliable and convenient. In the keys security aggregation process, each $S{M_i}$ negotiates the masks ${sk_{i,o}} = {x_i}s_o^{pk}(i \in \mathcal{U};o \in \mathcal{U},o \ne i )$ with all other SMs, and the mask agreement is based on the computational Difﬁe-Hellman hard problem. Suppose the SO wants to get the private key $s_i$ of the $S{M_i}$ after colluding with the MN, it still needs to collude with $m-2$ SMs, which is not achievable in practice. Therefore, our scheme resists collusion attacks.

\subsection{Effective defense evaluation}
In this sub-section, we calculate the probability of successful attacks by the attacker in two scenarios and illustrate the effectiveness of the scheme through mathematical proofs. 
\subsubsection{Scenario 1}Network attackers may destroy data before it is transmitted, during its transmission, and after it is received by the MN to render the system inoperable.
\subsubsection{Scenario 2}Network attackers may tamper with the original data before it is transmitted, during data transmission, and after it is received by the MN (before it is broadcast) to allow false data to be verified.

The attack methods and success probabilities of data being destroyed and tampered with before, during, or after transmission are summarized in Table VII.

\begin{table}[!t]
\renewcommand\arraystretch{2.3}
\caption{Probability of successful attacks\label{tab:table7}}
\centering
\begin{tabular}{|c|c|c|c|}
\hline
\multicolumn{2}{|c|}{Stages} & Scenario 1 & Scenario 2 \\
\cline{1-4}
\multirow{2}*{ \makecell{Pre-data\\transmission}} & \makecell{Attack \\method}  & Hack into $m$ SMs  & \makecell{Hack into $m$ SMs;\\ Get the keys} \\
\cline{2-4}
&  Probability  & $\prod\limits_{i = 1}^m {{P_{S{M_i}}}} $ & $\prod\limits_{i = 1}^m {{P_{S{M_i}}}}  \cdot \prod\limits_{i = 1}^m {{P_{{k_i}}}} $ \\
\hline
\multirow{2}*{\makecell{Data in \\ transit}} & \makecell{Attack\\ method} & Hack $m$ channels  & \makecell{Hack $m$ channels;\\ Get the keys}\\
\cline{2-4}
&  Probability  & $\prod\limits_{i = 1}^m {{P_{{C_i}}}} $   & $\prod\limits_{i = 1}^m {{P_{{C_i}}}}  \cdot \prod\limits_{i = 1}^m {{P_{{k_i}}}} $ \\
\hline
\multirow{2}*{\makecell{Data \\ received}} & \makecell{Attack\\ method}  & Hack into MN  & \makecell{Hack into $m$ SMs;\\ Get the keys} \\
\cline{2-4}
&  Probability  & ${P_{MN}}$   & $\prod\limits_{i = 1}^m {{P_{S{M_i}}}}  \cdot \prod\limits_{i = 1}^m {{P_{{k_i}}}} $ \\
\hline
\end{tabular}
\end{table}

For scenario 1, we suppose that the probability of an attacker hacking into a smart meter is denoted as ${P_{SM}}$, $0 < {P_{SM}} < 1$, and the probability of an attacker hacking into a channel is denoted as ${P_C}$, $0 < P_C < 1$. To make the system unworkable, the attacker needs to attack $m$ smart meters with success probability $\prod\nolimits_{i = 1}^m {{P_{S{M_i}}}} $ before data transmission, $m$ channels with success probability $\prod\nolimits_{i = 1}^m {{P_{{C_i}}}} $ during data transmission, and after the MN accepts the data, the success probability of the attack is ${P_{MN}}$. However, because $m$ is large, the attacker's probability of hacking into the smart meters is extremely low, and even if it is destroyed during the data transmission phase, it can still be detected from the data signature to discover and eliminate this attack, and meanwhile, when the MN is attacked and the data is destroyed, all other SMs will find the wrong data in the consensus phase and re-vote to select a new MN, so our scheme has good defense capability under scenario 1.

For scenario 2, we suppose that the probability of an attacker stealing the private key of the smart meter is denoted as ${P_k}$, $0 < P_k < 1$. When the attacker wants to tamper with the data in the smart grid, the probability of a successful attack before data transmission is $\prod\nolimits_{i = 1}^m {{P_{S{M_i}}}}  \cdot \prod\nolimits_{i = 1}^m {{P_{{k_i}}}} $, during data transmission is $\prod\nolimits_{i = 1}^m {{P_{{C_i}}}}  \cdot \prod\nolimits_{i = 1}^m {{P_{{k_i}}}} $, after the MN receives the data, the probability of a successful attack is $\prod\nolimits_{i = 1}^m {{P_{S{M_i}}}}  \cdot \prod\nolimits_{i = 1}^m {{P_{{k_i}}}} $. Compared with scenario 1, scenario 2 can be attacked with more demanding requirement conditions and a lower probability of a successful attack. From the above probabilistic analysis, it can be demonstrated that our scheme can perform the basic tasks in a more secure environment.

\subsection{System characteristic comparison}
The proposed scheme is compared with several other representative privacy-preserving electricity theft detection schemes for smart grid in terms of non-reliance on any trusted third party (TTP), data non-falsifiability (DNF), data non-repudiation (DNR), and data non-tamperability (DNT). As shown in Table VIII, the related work does not achieve all the desired characteristics of the smart grid, while only the proposed scheme achieves it. 

\begin{table}[!ht]
\caption{System characteristics comparison\label{tab:table8}}
\renewcommand\arraystretch{1.3}
\centering
\begin{tabular}{ccccc}
\hline
  & No TPP & DNF & DNR  & DNT \\
\hline
Joker et al.\cite{jokar2015electricity} &  Yes & No & No & No \\
Yao et al.\cite{yao2019energy} & No & Yes &  Yes & No\\
I.Ibrahem et al.\cite{ibrahem2020efficient} & No & No &  No & No\\
Richardson et al.\cite{richardson2016privacy} & No & No &  No & No\\
Our Scheme & Yes & Yes & Yes & Yes \\

\hline
\end{tabular}
\end{table}

\section{Conclusion}
In this paper, we propose a more secure blockchain-based privacy-preserving electricity theft detection scheme. The proposed scheme does not require a third party, which avoids the security and privacy issues brought about by a third party. Meanwhile, the distributed storage scheme of blockchain prevents security issues such as data tampering and forgery. In addition, a real dataset and environment are used for simulation evaluation. The experimental results show that the proposed scheme can detect malicious consumers more accurately with acceptable communication and computational overhead. System analysis shows that the proposed scheme is more secure compared to existing schemes. For our future work, we intend to improve the proposed scheme by reducing communication and computation overhead.

%\section*{Acknowledgments}

\bibliographystyle{IEEEtran}
\bibliography{ref}

%\section{Biography Section}
%If you have an EPS/PDF photo (graphicx package needed), extra braces are
% needed around the contents of the optional argument to biography to prevent
% the LaTeX parser from getting confused when it sees the complicated
% $\backslash${\tt{includegraphics}} command within an optional argument. (You can create
% your own custom macro containing the $\backslash${\tt{includegraphics}} command to make things
% simpler here.)
 
\vspace{11pt}

\end{document}